\documentstyle[12pt]{article}
\newcommand{\bq}{\begin{eqnarray}}
\newcommand{\eq}{\end{eqnarray}}
\newcommand{\ra}{\rightarrow}
\newcommand{\th}{\tilde\theta}
\newcommand{\ftt}{\frac{\th}{N_c}}
\newcommand{\ov}{\overline}

\begin{document}

\begin{flushright}
{\bf BINP 98-61}\\
(revised)\\
hep-th/9808092\\
\end{flushright}

\begin{center} {\bf PROPERTIES OF THE ${\bf SU(N_c)}$ YANG-MILLS
 VACUUM STATE}\\

\vspace{.5cm}

{\bf Victor Chernyak}\footnote{e-mail: chernyak@inp.nsk.su}\\

\vspace{.5cm}

 {Budker Institute of Nuclear Physics, 630090 Novosibirsk-90, Russia}
\end{center}

\vspace{.5cm}

\begin{abstract}
The asymptotic behaviour of the vacuum energy density, ${\ov E}_{vac}
(\theta)$, at $\theta\ra\pm\, i\,\infty$ is found out. A new interpretation 
and a qualitative discussion of the ${\ov E}_{vac}(\theta)$ behaviour 
are presented. It is emphasized that the vacuum is doubly degenerate at
$\theta=\pi$, and the quark electric string can terminate on the domain wall 
interpolating between these two vacua. The potential of the monopole field 
condensing in the Yang-Mills vacuum is obtained.
\end{abstract}

\vspace{1cm}

\hspace*{0.5cm}{\bf 1.}\hspace{.5cm}
Let us consider the (Euclidean) YM $SU(N_c)$ theory defined by the integral:
\bq  
Z=\sum_k\int dA_\mu\,\delta(Q-k)\exp \left\{-\int dx\,I_o(x)\right\},\,\,
\quad I_o=\frac{1}{4g_o^2}\,G_{\mu\nu}^2\,,															
\eq
where Q is the topological charge, and let us integrate out the gluon fields 
with the constraint that the composite field $G_{\mu\nu}^2$ is fixed: 
\footnote
{ This is what differs our approach from those proposed by G.
Savvidy \cite{Sav},\, the latter one leading to instabilities.}
$$
Z=\int dS\,\theta(S)\exp\left\{-\int dx\, I(S)\right\},\quad 
I(S)=N_c^2 \frac{b_o}{4}\left [\,S\ln\frac{M_o^4}{{\bar \Lambda}^4}
+f(S,M_o^2)\right]\,, 
$$
\bq 
\exp \left\{-\int dx\,N_c^2 \frac{b_o}{4} f(S,M_o^2)\right\}=
\sum_k\int dA_\mu\,\delta(Q-k)\,\,\,\delta\left(S-\frac{G^2_{\mu\nu}}
{32\pi^2 N_c}\right)\,, 
\eq
where $M_o$ is the ultraviolet cut off,\, and we have substituted: $b_o=
11/3,\,1/g_o^2=(\,N_c\, b_o/32\pi^2\,)\ln\left(M_o^4/
{{\bar\Lambda}^4}\right).$

Let us emphasize that the function $f(S,M_o^2)$ does not know 
about $\bar\Lambda$. Thus, recalling for a 
renormalizability of the theory,\, it can be written in the form:
$$
f(S,M_o^2)=-S\ln \left ( \frac{M_o^4}{S}\right )+S\,C_N+
\frac{1}{S^{3/2}}\partial_{\mu} S\,\Delta_o \left (\frac{q^2=-\partial^2}
{S^{1/2}} \right )\partial_\mu S+
$$
\bq
+O\left ( (\partial_{\mu}S)^4\right )+\cdots+\frac{1}{N_c^2} f_{s}
(S,M_o^2)\,.
\eq
So,\, the partition function takes the form:
$$ 
Z=\int dS\,\,\theta(S) \exp \left \{  -\int dx\, N_c^2 \frac{b_o}{4} \left
[ L_o(S)+S\,C_N+\frac{1}{N_c^2}f_{s}( S,M_o^2 )\,\, \right ]  \right \}\,,
$$  
\bq L_o(S)=S\,\ln \left ( \frac{S}{{\bar\Lambda}^4}\right )
+\frac{1}{S^{3/2}}\partial_{\mu} S\,\Delta_o \left (\frac{q^2}
{S^{1/2}} \right )\partial_\mu S+O\left ( (\partial_{\mu}S)^4\right )+\cdots 
\eq

Because we have not integrated yet the quantum loop contributions 
of the field S, 
the action still depends (logariphmically) on the cut off $M_o^2$ (through 
the function $f_s$ in eq.(4)), and this dependence will disappear 
only after the quantum loop corrections of the fields S will be 
accounted for completely. However, these quantum corrections are amenable to 
the standard $1/N_c$ - counting. And because the field S is only one out of
$\sim N_c^2$ degrees of freedom, the $1/N_c$ - counting implies that its
quantum fluctuations will give only $1/N_c^2$ - corrections to the action. 
\footnote{
In all the above formulae the leading $N_c$ -dependence is written out
explicitly, while the dependence on the nonleading corrections in powers
of $1/N_c$ is implicit.
}
This means, in particular, that if we neglect the quantum loop contributions 
of the field S altogether, then the above action in eq.(4) will be a 
generating functional of the 1-PI (one particle irreducible) Green functions
of the field S, and will differ from the exact generating functional by
the $1/N_c^2$ corrections only. In particular, the potential 
(i.e. without space-time derivatives) term will be: $U= N_c^2 \frac{b_o}
{4} S\, [ \,\ln \left ( \frac{S}{{\bar\Lambda}^4}\right )+C_N\, ] \,,$ etc.

Let us emphasize now that, on the one hand, the constant $C_N$ in eq.(4) can 
be considered as the external (constant) source of the field S and, on the
other hand, all the dependence on this source can be absorbed into a
redefinition of the scale parameter ${\bar\Lambda}$. For our further 
purposes, let us substitute it (temporarily) by the local source, $C_N(x).$

Now, accounting for the quantum loop contributions of the field S and 
considering $C_N(x)$ as a source, the partition function can be written in 
the form of the Legendre fransform:
\bq
Z=\oint d S\, \theta (S)\,\exp\left \{-\int dx N_c^2 \frac{b_o}{4} \left [
L_o(S)+\frac{1}{N_c^2}\,\delta L(S)+S\,C_N(x)\right ] \right \}\,,
\eq
where the integral $\oint dS$ in eq.(5) means that 
quantum loop corrections are ignored now, so that $L=
(\,L_o+(1/N_c^2)\,\delta L\,)$ is the exact generating functional of the 
1-PI Green functions of the field S. Finally, using the above
described property that all the dependence on the source $C_N(x)$ (when it
becomes constant) can be absorbed into a redefinition of ${\bar\Lambda}$, 
its potential part becomes completely fixed, and so $L$ can be written in 
the form ($B_N=1+O(1/N_c^2)$ is another constant):
\bq
L=\left [\,S\ln\left(\frac{S}{B_N{\bar\Lambda}^4}\right)+\frac{1}{S^{3/2}}
\partial_{\mu} S\,\Delta_{1}\left (\frac{q^2}{S^{1/2}},N_c \right )
\partial_\mu S+O\left( (\partial_{\mu}S)^4\right ) +\cdots \right ]\,
\eq
where, for instance, the function $\Delta_1$ approaches $\Delta_o$ at $N_c
\gg 1$, etc.

Let us emphasize that, as it is seen from its derivation, the action $L(S)$ 
in eqs.(5,\,6) is not some approximate effective or low energy action but the 
exact one,\, and the exact answers for all correlators of the field S can be 
obtained from this action calculating the tree diagrams only (as all quantum 
loop contributions of the field S were accounted for already).
\footnote{
To simplify, we used the one-loop $\beta$-function 
in the above equations, neglecting the terms with the nonleading dependence
on the cut off, $M_o$. More precisely,\, 
to obtain the renormalization group invariant expressions we have to fix the 
renormalization group invariant quantities,\, e.g. $(-1/N_c^2b_o)T_{\mu\mu}= 
z_G(g_o^2)S=(1+O(g_o^2))S$ in this case, where $T_{\mu\mu}$ is the 
energy-momentum tensor trace. Having this in mind,\, it is implied everywhere
below that appropriate renormalization factors,\, $z_i$,\, are included 
already into definitions of all fields and parameters we deal with. }

The above described way of obtaining the exact generating functional of the
1-PI Green functions of composite fields will be widely used below and in 
the subsequent paper \cite{II}. In a few words, it can be formulated as 
follows. Let us suppose that we have some quantum field $\phi$ (elementary
or composite) with the (Euclidean) partition function:
\bq
Z=\int d\phi \,\, \exp \left \{-\int dx\,\left [\, L_o(\phi)-J(x)\phi \,\,
 \right ] \right \}\,, 
\eq
where $J(x)$ is a source. After integrating out all quantum loop effects, it
can be represented in the form of the Legendre transform:
\bq
Z=\oint d\phi \exp \left \{-\int dx\,\left [\, L_o(\phi)+\Delta L(\phi)
-J(x)\phi\, \right ] \right \}\,,
\eq
where $L(\phi)=L_o(\phi)+\Delta L(\phi)$ is the exact generating functional
of the 1-PI Green functions of the field $\phi$. I.e., the exact correlators
of the field $\phi$ are obtained from Z in eq.(8) as follows. To obtain
the n-point Green function:\\ 
a) decompose the field $\phi$ as: $\phi(x)=\phi_{class}(x)+\phi_{quant}(x)$, 
where $\phi_{class}(x)$ is determined from the stationary point condition:
$[\,\delta L(\phi)/\delta \phi\,]_{\phi=\phi_{class}}=J(x)$\,;\\ 
b) decompose $L(\phi_{class}+\phi_{quant})$ in powers of $\phi_{quant}$\,;\\
c) the normalization of $\oint$ is such that: $\oint d\phi_{quant} \exp \left 
\{-\int dx L_2 \right \}=1$, where $L_2$ is the quadratic in $\phi_{quant}$ 
part of $L$. Other loop contributions 
are always ignored in $\oint$, so that: $Z=\exp \{-{\ov E}_{vac}\},\,
{\ov E}_{vac}=L(\phi_{class})-J\phi_{class}$\,;\\ 
d) put the factor $Z^{-1}
\phi_{quant}(x_1)\dots \phi_{quant}(x_n)$ inside the integral in eq.(8); \\
e) calculate the integral over $\phi_{quant}(x)$ keeping the tree diagrams 
only.   

We would like to emphasize here that, clearly, the function
$L(\phi)$ is independent of the source $J(x)$. So, we can freely
change the source or even to introduce it afterwards, - $L(\phi)$ (considered
as a function of $\phi$ and its derivatives) will stay
intact, and only $\phi_{class}$ will change when changing the source. 
And we will widely use this property below and in \cite{II}.   

Little is known about the  inverse propagator of
the S-field,\, except what can be obtained from the asymptotic freedom and
operator expansions. 
Let us define the propagator: $i\int dx\exp\,(iqx)\langle S(x)S(0) \rangle
_{con}=D(q^2,\,\Lambda^2)$. At $q^2\gg \Lambda^2$:
$$
D(q^2,\,\Lambda^2)= \left \{ \left [ \,C_1\,\frac{q^4}{\ln(q^2/\Lambda^2)}
\left (\, 1+O(\alpha_s\,) \right )\right ]
+\left [ C_2\,\frac{\Lambda^4}{\ln(q^2/\Lambda^2)}
\left (\,1+O(\alpha_s)\right ) \right ] \right.+
$$
$$
\left. +\left [ C_3\,\alpha_s^{\delta}(\frac{q^2}{\Lambda^2})\frac{\Lambda^6}
{q^2}\left (\,1+O(\alpha_s)\right )\right ]+\cdots \right \}.
$$
Clearly, the behaviour of $\Delta_{1}(q^2/S^{1/2},N_c)$ in eq.(6) at $q^2
\gg S^{1/2}$ can be reconstructed from the above behaviour of D, with a 
replacement: $\Lambda^2\ra [eS]^{1/2}$.
\footnote{
As can be easily checked, the behaviour of $\Delta_{o}(q^2/S^{1/2})$ at large 
$q^2$ agrees precisely with that the quantum loop corrections of the field S 
give only $\sim (1/N_c)\ln M_o^2$ contribution to the renormalization of the 
charge $1/g_o^2$, i.e. a relative $\sim 1/N_c^2$ correction.  
}

In what follows we will be interested
mainly in the potential part (i.e. without space-time derivatives) of
the action and will ignore all terms with such derivatives.

It follows from eqs.(5,6) that the field S condenses in the YM-vacuum,\,
and the values of the condensate and the vacuum energy density are:
\bq 
{\ov S}=e^{-1}\,\Lambda^4\,,\quad {\ov E}_{vac}=-N_c^2\frac{b_o}{4}\,
\Lambda^4\,<\,0\,.
\eq

As it is,\, the potential in eq.(6) is well known (see e.g. \cite
{Shifm}). However,\, the field S was considered previously mainly as
the effective dilaton field,\, i.e. the interpolating field of the
lightest scalar gluonium. Besides, such Lagrangeans were usually considered 
as some "effective" Lagrangeans, the meaning of "effective" was obscure,
as well as their connection with the original YM-Lagrangean. Our approach
allows to elucidate its real origin and meaning and,\, on this basis,\, 
to use it for the investigation of the vacuum and correlators properties. 

\hspace*{.5cm}{\bf 2.}\hspace{.5cm}
Let us extend now our theory and add the $\theta$-term to the action
$I_o$ in eq.(1),\, and let us integrate out the gluon fields with
the fields  S(x) and P(x) both fixed ($\th=i\theta,\, Q=N_c\int dx P)$:
$$
Z=\sum_k\int dS\,\theta(S)\,\int dP\,\theta(S-|P|)\,\delta(Q-k)\times
$$
\bq
\times\exp\left\{-\int dx N_c^2\left [\frac{b_o}{4}S 
\ln \frac{M_o^4}{\Lambda^4}+
S\,I_o(\frac{P}{S},\frac{M_o^4}{S}) -\ftt P\right ] \right\}\,,
\eq
$$ 
\exp\left\{-\int dx N_c^2\,S\,I_o(\frac{P}{S},\frac{M_o^4}{S})\right\}=
\int dA_\mu \delta\left(S-\frac{G^2_
{\mu\nu}}{32\pi^2N_c}\right) \delta\left(P-\frac{G{\tilde G}}{32\pi^2N_c}
\right).
$$

The positivity of the Euclidean integration measure (at real positive $\th$)  
leads to a number of useful sign inequalities, like ${\overline S}
(\th) \geq {\overline P}(\th)\geq 0,$ etc. Because ${\overline P}
(\th)=(b_o/4)\,d{\overline S}(\th)/d\th$, this shows that ${\overline S}
(\th)$ grows monotonically with $\th$ (so that the energy density decreases 
monotonically with $\th$).
 
Because the $\theta$ - term in the Lagrangean can be considered as a source
of the field P (moreover, we can even replace it by the local function, 
$\theta(x)$), integrating out the quantum loop contributions of the S and P
fields and using the above described considerations, the partition
function can be rewritten in the form:
$$
Z=\sum_k\oint dS\,\theta(S)\,\oint dP\,\theta(S-|P|)\,\delta(Q-k) \times
$$
\bq
 \times \exp\left\{-\int dx N_c^2\left[I(S,P)-\ftt P\right]\right\}\,,
\eq

Let us write the general form of the potential in eq.(11):
\bq
U(S,P)=\frac{b_o}{4}S\left[\,\ln\left(\frac{S}{\Lambda^4}\right)+
f(z=\frac{P}{S})\, \right]-\ftt P\,,
\eq
and the stationary point equations:
\bq 
\ftt=\frac{b_o}{4}\,f^{\prime}(\bar z),\quad \frac{b_o}{4}\left[\ln \left
(\frac{e\ov S}{\Lambda^4}\right)+f(\bar z)\right]=\ftt \bar z.
\eq
Consider now the behaviour at $\th\ra\infty$. It is seen from eq.(13)
that $\bar z(\th)$ approach $z_o$,\, - the singularity point of $f^{\prime}
(z)$. It looks physically unacceptable if 
$f(z)$ had singularities (say, poles 
or branch points) inside the physical region $0<z<1.$ They can develop at the
edges of the physical region only: $z\ra 0$ or $z\ra 1$. The behaviour of
$f(z)$ at $z\ra 0$ is regular however,\, $\sim z^2$. So,\, we conclude that:
${\bar z}(\th)={\ov P}(\th)/{\ov S}(\th)\ra 1$ at $\th\ra \infty$. 
Supposing that the singularity is gentle enough so that $f(1)$ is finite,
one obtains then from eq.(13):
\bq
{\ov P}(\th)\ra{\ov S}(\th)\ra C\exp\left\{\frac{n}{b_o}\ftt\right\},\,\,
C=e^{-f(1)-1}\Lambda^4,\,\,\,\,\, \th\ra\infty,
\eq
where $n=4$ is the space-time dimension (compare with the $CP^N$ model,\, 
see appendix).

To illustrate the typical properties, let us consider the simple model 
for $U(S,P)$ in eq.(12). It looks (in Minkowsky space) as follows:
\bq 
 U(S,P)=\frac{1}{2}\left\{(S+iP)\ln\left(\frac{S+iP}{\Lambda^4
e^{i\theta/N_c}}\right)+h.c.
\right\}-\frac{1}{12}S\ln\left(\frac{S}{\Lambda^4}\right).
\eq
When integrated over S and P fields it gives (at the space-time volume
$V\ra \infty$):
$$
{\ov S}(\theta)=e^{-1}\Lambda^4\left[\cos^{4/b_o}\left(\frac{\theta}{N_c}
\right)\right]_{2\pi}\,, 
$$
\bq
{\ov E}_{vac} =-N_c^2\frac{b_o}{4}\,{\ov S}
(\theta)\,, \quad \ov P(\theta)=-\frac{1}{N_c}\frac{d {\ov E}_{vac}(\theta)}
{d\theta}\,\,.
\eq
Here the notation $[f(\theta/N_c)]_{2\pi}$ means that this function is
$f(\theta/N_c)$ at $-\pi\leq \theta \leq \pi$,\, and is glued then to be
periodic in $\theta\ra \theta+2\pi k$,\, i.e.:
\bq
\left [ f(\frac{\theta}{N_c})\right]_{2\pi}=
\min_k\,\, f\left(\frac{\theta+2\pi k}
{N_c}\right)\,.
\eq
Evidently,\, the above periodically glued structure of the vacuum energy
density results from the quantization of the topological charge in our
theory,\, and is not connected with a concrete form of the action
$I(S,P)$ in eq.(11). 

The above model does not pretend to be the exact answer, but it is 
simple, has a reasonable qualitative behaviour in the whole complex plane
of $\theta$, and obeys the right asymptotic behaviour at $\theta\ra 
\pm i\infty$. 

It was inspired by a "different status" of two
parts in $b_o/4=(1-\frac{1}{12})$. The first part is connected with zero
mode contributions in the instanton background, and the analytic in
$\chi=(S+iP)$ term in eq.(15) is expected to be connected with this
part of $b_o$ only, while the second, nonanalytic in $\chi$, part of
eq.(15) is expected to be due to the "nonzero mode part" of $b_o$ only.

In Euclidean space, one has from eq.(16):
\bq
{\ov S}(\th)=e^{-1}\Lambda^4\,\cosh^{4/b_o}\left(\ftt\right),\quad {\ov P}
(\th)=\tanh \left(\ftt \right){\ov S}(\th)\,,
\eq
so that the asymptotic behaviour, eq.(14), is reproduced, and ${\bar z}
(\th)={\ov P}/{\ov S}\ra 1$ at large $\th$. In principle, the equation (18) 
can be checked in lattice calculations.
\footnote{
There is another simple model: $(1/N_c^2)U=\frac{1}{2}\left \{
\phi\ln\frac{\phi}{\Lambda^4e^{i\theta/N_c}}+h.c. \right \},\,\, 
\phi=(b_o/4)S+iP\,,$
leading to ${\ov E}_{vac}(\theta)\sim [\cos (\frac{b_o}{4}\frac
{\theta}{N_c})]_{2\pi}$, i.e.
with the same asymptotic behaviour at $\theta\ra \pm i\infty$. The problem
with it is that there are reasons to expect that the vacuum energy is
exactly zero at $\theta=\pi$ and $N_c=2$, and this model does not fulfil
this, while eq.(15) does. Besides, we see no reasons here for the potential 
to be analytic function of one variable.}  

One point is worth mentioning in connection with the asymptotic behaviour,
eq.(14). Based on the
quasiclassical (one loop) instanton calculations \cite{ins}, one would
expect the following qualitative picture. In $SU(N_c)$, each instanton
splits up into $N_c$ "instantonic quarks" which appear as appropriate
degrees of freedom in the dense instanton ensemble. As a result, each
instantonic quark carries the factor $\exp\{i\theta/N_c\}$ in its density.
So, one would expect the behaviour ${\bar P}(\th)\ra {\bar S}(\th)\sim
\exp\{{\tilde \theta}/N_c\}$ at large $\th$, in disagreement with eq.(14). 
We conclude that something is missing here in the above picture of 
instantonic quarks, even in its $\theta$\,-dependence.

\hspace*{.5cm}{\bf 3.}\hspace{.5cm}
We will describe now a new qualitative interpretation of (glued) 
periodicity properties of the vacuum energy density, ${\ov E}_{vac}(\theta)$. Let 
us consider first $N_c=2$ for simplicity, and let us suppose the "standard"
picture of the confinement mechanism to be valid. I.e., the internal
(dynamical) Higgs breaking $SU(2)\ra U(1)$ takes place and, besides, the
$U(1)$ magnetic monopoles condense. As shown by E. Witten \cite{Witten}, 
the monopoles turn into the dyons with the $U(1)$ electric charge $ \theta/
2\pi$ when the $\theta$ - term is introduced. Although, strictly speaking,
the Witten result was obtained in the quasiclassical region only, because
the effect is of a qualitative nature there are all the reasons to expect 
it will survive in the strong coupling region also ( and only the units of 
the electric and magnetic charges will change).

So, there are pure monopoles and antimonopoles with the magnetic and
electric charges $(g,\,e)=(1,\,0)$ and $({\bar g},\,{\bar e})=(-1,\,0)$ 
in the condensate at $\theta=0$, while they turn into the dyons and antidyons 
with the charges $d_1^{\theta}=(1,\,\theta/2\pi)$ and $\bar {d_1^{\theta}}=
(-1,\,-\theta/2\pi)$ respectively, as $\theta$ starts to deviate from zero. 
As a result, the vacuum energy density begins to encrease, as it follows 
from general considerations.

It is a specific property of our system that there are two types of
condensates made of the dyons and antidyons with the charges: 
$\{(1,1/2);\,(-1,-1/2)
\}$ and $\{(1,-1/2);\,(-1,1/2)\}$, and having the same energy density. 
\footnote{
This can be seen, for instance, as follows. Let us start from the pure
monopole condensate at $\theta=0$ and let us move along the path: 
$\theta=0\rightarrow \theta=\pi$. The vacuum state will consist of
$d_1 (1,\frac{1}{2})$ - dyons and ${\bar d}_1 (-1,-\frac{1}{2})$ -
antidyons. Let us move now along the path: $\theta=0\rightarrow 
\theta=-\pi$. The vacuum state will consist now of
$d_2 (1,-\frac{1}{2})$ - dyons and ${\bar d}_2 (-1,\frac{1}{2})$ -
antidyons. Because the vacuum energy density is even under $\theta\ra
-\theta$, these two vacuum states are degenerate.

Let us emphasize that the existence of two vacuum states at
$\theta=\pi$ does not follow from the symmetry considerations alone (like
${\ov E}_{vac}(\theta)={\ov E}_{vac}(-\theta)$ and ${\ov E}_{vac}(\theta)=
{\ov E}_{vac}(\theta 
+ 2\pi k)$). It is sufficient to give a counterexample. So, let us consider
the Georgy-Glashow model, with the large Higgs vacuum condensate resulting
in $SU(2)\ra U(1)$. In this case, the $\theta$\,-dependence of the vacuum 
energy density is due to a rare quasiclassical gas of instantons, and is
$\sim \cos(\theta)$. All the above symmetry properties are fulfilled, but
there is only one vacuum state at $\theta=\pi$.
}
Besides, these two states belong to the same world as they are reachable one 
from another through a barrier, because there are electrically charged 
gluons, $(0,\pm 1)$, in the spectrum which can recharge these $(1,\pm 1/2)$ 
- dyons into each other. In contrast, the two vacuum states, $|\theta 
\rangle $ and $|-\theta \rangle $ at $\theta \neq 0,\, \pi$ are 
unreachable one from another and belong to different worlds, as there is no
particles in the spectrum capable to recharge the $(1,\pm \theta/2\pi)$ -
dyons into each other. 

Thus, the vacuum becomes twice degenerate at $\theta= \pi$, so that the 
"level crossing" (in the form of rechargement: $\{d_1=(1,1/2),\,{\bar d}_1=
(-1,-1/2)\}\ra \{d_2=(1,-1/2),\,{\bar d}_2=(-1,1/2)\}$)
can take place if this will lower the energy density.
And indeed it lowers, and this leads to a casp in ${\ov E}_{vac}(\theta)$. At 
$\theta>\pi$, the vacuum is filled now with new dyons with the charges: $d_
2^{\theta}=(1,\,-1+\theta/2\pi),\, {\bar d}_2^{\theta}=(-1,\,1-\theta/2\pi)$. 
As $\theta$ increases further, the electric charge of the $d_2^{\theta}$\,
-dyons decreases, and the vacuum energy density decreases with it. Finally, 
at $\theta=2\pi$, the $d_2^{\theta}$\,-dyons (which were the (1,\,-1)-dyons 
at $\theta=0$) become pure monopoles, and the vacuum state becomes 
exactly as it was at $\theta=0$, i.e. the same condensate of the pure 
monopoles and antimonopoles.

We empasize that, as it follows from the above picture, it is wrong to
imagine the vacuum state at $\theta=2\pi$ as, for instance, a condensate of 
the dyons with the charges (1,\,-1), degenerate in energy with the pure 
monopole condensate at $\theta=0$.
\footnote{
In this respect, the widely used terminology naming the two singularity points,
$u=\pm \Lambda^2$, on the ${\bf{\cal{N}}}=2\,\,\, SU(N_c=2)$ SYM moduli space
as those where monopoles and respectively dyons become massless, is not
quite adequate. Indeed, let us start from the vacuum $u=\Lambda^2$ where, by
definition, the massless particles are pure monopoles, and
let us move, for instance, along a circle to the point $u=-\Lambda^2$. On
the way, the former massless monopole increases its mass because it becomes 
the $d_1^{\theta}=(1,\,\theta/2\pi)$ - dyon, while the former massive  
$d_2^o=(1,\,-1)$ - dyon diminishes its mass as it becomes the $d_2^{\theta}=
(1,\,-1+\theta/2\pi)$ - dyon. When we reach the point $u=-\Lambda^2,$ i.e.
$\theta=2\pi$, the former dyon becomes massless just because it becomes the 
pure monopole here. So, an observer living in the world with $u=-\Lambda^2$ 
will also see the massless monopoles (not dyons, and this is distinguishable
by their Coulomb interactions), as those living in
the world with $u=\Lambda^2$.
}

As for $SU(N_c>2)$, the above described picture goes through without
changes if we suppose that $SU(2)\ra U(1)$ is replaced by $SU(N_c)\ra U(1)^
{N_c-1}$, and there are $(N_c-1)$ types of monopoles in the condensate, with 
the magnetic charges $(m^A)_i=\pm n_i^A,\, A=1,...,N_c-1,\,i=1,...,N_c.$ (For 
instance, it has been proposed by G.'t Hooft \cite{Hooft} 
that the good candidates 
to condense are the "minimal monopoles" having two consecutive and opposite 
charges, i.e. $n_i^A=\delta_i^A-\delta_{i+1}^A$ in this case). Now, when the 
$\theta$- term is introduced into the Lagrangean, each one of these $(N_c-1)$
monopoles will turn into the dyon with the same type electric charge: $(e^A)_
i=\pm (\theta/2\pi)\,n^A_i$. So, the above described picture of dyon
rechargement will be applicable as well, resulting in a cusp in ${\ov E}_
{vac}(\theta)$ at $\theta=\pi$. 

In fact, for the above described mechanism to be operative, 
there is no need to trace the real dynamical picture underlying a cusped
behaviour of ${\ov E}_{vac}(\theta)$ in the infinitesimal vicinity of 
$\theta=\pi$. Formally, it is sufficient to say: 
"there is a possibility for electrically charged degrees of freedom 
to rearrange themselves without changing the volume energy". However, it may 
be useful to have a more visible picture of how it can proceed. At least 
formally, it can be thought as a typical first order phase transition. 
In a space with the coherent condensate of the $d_1=(1,1/2)$ - dyons and 
${\bar d}_1=(-1,-1/2)$ - antidyons, there appears a critical bubble with the 
coherent condensate of the $d_2=(1,-1/2)$ - dyons and ${\bar d}_2=(-1,1/2)$ -
antidyons deep inside, and with a transition region surface (domain wall) 
through which the averaged densities of two type dyons interpolate smoothly. 
This bubble expands then over all the space through a rechargement process 
$d_1+{\bar d}_1\ra d_2+{\bar d}_2$ occuring on a surface. This rechargement 
can also be thought as going
through a copious "production" of charged gluon pairs, so that the
underlying processes will be: $[d_1=(1,1/2)]+[{\bar g}=(0,-1)]\ra [d_2=(1,
-1/2)]$ and $[{\bar d}_1=(-1,-1/2)]+[g=(0,1)]\ra [{\bar d}_2=(-1,1/2)]$.

Some analogy with the simplest Schwinger model may be useful at this point, 
in connection with the above described rechargement process.
Let us consider first the pure $QED_2$ without finite mass charged 
particles, and let us put two infinitely heavy "quarks" with the charges
$\pm \theta/2\pi$ (in units of some $e_o$) at the edges of our
(infinite length) space. It
is well known that this is equivalent to introducing the $\theta$\,-angle
into the $QED_2$ Lagrangean. As a result, there is the empty vacuum at
$\theta=0$, and the long range Coulomb "string" at $\theta \neq 0$. The
vacuum energy density increases as: $E(\theta)=C_o e_o^2\, \theta^2,\, 
C_o=const,$ at any $0\leq \theta < \infty$.

Let us add now some finite mass, $m\gg e_o$, and of unit charge $e_o$
field $\psi$ to the Lagrangean. When there are no external charges, this
massive charged field can be integrated out, resulting in a small
charge renormalization. But when the above "quarks" are introduced, the
behaviour of $E(\theta)$ becomes nontrivial: 
$E(\theta)=C_o e_o^2 \min_k (\theta+2\pi k)^2$. So, $E(\theta)=
C_o e_o^2\, \theta^2$ at $0\leq \theta \leq \pi$, and
$E(\theta)=C_o e_o^2\, (2\pi-\theta)^2)$ at $\pi\leq \theta \leq 2\pi$.

The reason is clear. The external "quark" charge becomes equal 1/2 at $\theta
=\pi$. At this point, a pair of $\psi$\,- particles is produced from the
vacuum, and they separate so that to recharge the external "quarks": 
$\pm 1/2 \ra  \mp 1/2$. As a result of this rechargement, there appears a 
cusp in $E(\theta),$ and $E(\theta)$ begins to decrease at $\theta> \pi$, 
so that the former "empty" vacuum is reached at $\theta=2\pi$.

Let us return however to our dyons. The above described picture predicts 
also a definite behaviour of the topological 
charge density, ${\ov P}(\theta).$  
At $0<\theta<\pi$, i.e. in the condensate of the $d_1^{\theta}=(1,\theta/
2\pi)$ - dyons and ${\bar d}_1^{\theta}=(-1,-\theta/2\pi)$ - antidyons, the
product of signs of the magnetic and electric charges is positive for both 
$d_1^{\theta}$ - dyons and ${\bar d}_1^{\theta}$ - antidyons. Thus, these
charges give the correlated field strengths:
${\vec E}\,\Vert \,\theta {\vec H},\, {\vec E \cdot \vec H} > 0$, and both
species contribute a positive amount to the mean 
value of the topological charge density, so that ${\ov P}_1(\theta)>0$ and 
grows monotonically with $\theta$ in this interval (following increasing 
electric charge, $\theta/2\pi$, of the dyon). 
\footnote{
The case $N_c=2$ may be an exception, see footnote 6 and eq.(16), 
and also below.}

On the other side, at $\pi<\theta<2\pi$, i.e.
in the condensate of the $d_2^{\theta}=(1,-1+\theta/2\pi)$ - dyons and 
${\bar d}_2^{\theta}=(-1,1-\theta/2\pi)$ - antidyons, the
product of signs of the magnetic and electric charges is negative for both 
$d_2^{\theta}$ - dyons and ${\bar d}_2^{\theta}$ - antidyons. Thus,
both species 
contribute a negative amount to ${\ov P}_2(\theta)$, such that: ${\ov P}_2
(\theta)=-{\ov P}_1 (2\pi-\theta)$, and ${\ov P}(\theta)$ jumps reversing 
its sign at $\theta=\pi$ due to a rechargement. 

Clearly, at $0\leq\theta<\pi$, the condensate made of only the $d_1^{\theta}
=(1,\,\theta/2\pi)$ - dyons (recalling also for a possible charged gluon 
pair production) can screen the same type $d_k^{\theta}=
[\,const\,(1,\,\theta/2\pi)+(0,\,k)\,]$ - test dyon only $(k=0,\pm 1,
\pm 2, \dots$; and the same for the $d_2^{\theta}$ - dyons at $\pi< \theta
\leq 2\pi)$. So, the heavy quark-antiquark pair will be confined at 
$\theta\neq \pi$.

New nontrivial phenomena arise at $\theta=\pi$. Because there are two
degenerate states, i.e. the condensates of $(1,\pm \frac{1}{2})$ - dyons 
(and antidyons), a "mixed phase" configuration becomes possible with, for 
instance, each condensate filling a half of space only, and with the domain 
wall interpolating between them. This domain wall represents "a smeared
rechargement", i.e. a smeared over space interpolation of electrically
charged degrees of freedom between their corresponding vacuums, resulting
in a smooth variation of the averaged densities of both type 
dyons through the domain wall. Surprisingly, there is no confinement inside 
the bulk of such domain wall. 

The reason is as follows. Let us take the domain wall interpolating along 
the z-axis, so that at $z\ra -\infty$ there is a large coherent density of 
$d_1$-dyons, and at $z\ra \infty$, - that of $d_2$-dyons. 
As we move from the far
left to the right, the density of $d_1$-dyons decreases and there appears
also a small but increasing (incoherent) density of $d_2$-dyons. This small 
amount of $d_2$-dyons is "harmless", in the sense that its presence does 
not result in the screening of the corresponding charge. The reason is clear: 
the large coherent density of $d_1$-dyons keeps the $d_2$-dyons on the
confinement, so that they can not move freely and appear only in the form of
the rare and tightly connected neutral pairs, ${\bar d}_2d_2$, fluctuating
independently of each other. As we further move to the right, the 
$d_2$-dyons move more and more freely and their density increases, although 
they are still on the confinement. Finally, at some distance from the 
centre of the wall the "percolation" takes place, i.e. the $d_2$-dyons form 
a continuous coherent network and become released, so that the individual 
$d_2$-dyon can travel to arbitrary large distances (within its network). It 
should be emphasized that in this percolated region the coherent network of 
$d_1$-dyons still survives, so that these two networks coexist in the space. 

This system shares some features in common with the mixed state of the 
type-II superconductor in the external magnetic field. The crucial difference
is that the magnetic flux is sourceless inside the superconductor, while in
the above described system there are real charges (and anticharges) inside 
each network. So, polarizing itself appropriately, this system of charges is 
capable to screen any test charge.

As we move further to the right, the density of $d_2$-dyons 
continue to increase while those of $d_1$ continue to decrease. Finally, 
at the symmetric distance to the right of the wall centre the "inverse
percolation" takes place, i.e. the coherent network of $d_1$-dyons decays
into separate independently fluctuating droplets whose average density (and 
size) continue to decrease with increasing z. Clearly, the picture on the 
right side repeats in a symmetric way those on the left one, with the $d_1$ 
and $d_2$ dyons interchanging their roles.

Let us consider now the heavy test quark put inside the bulk of the domain 
wall, i.e. inside the "percolated" region. This region has the properties 
of the "double Higgs phase". Indeed, because the charges of two dyons,
$(1,\,1/2)$ and $(1,\,-1/2)$, are linearly independent, polarizing itself 
appropriately this mixture of the dyon condensate networks will screen any 
external charge put inside, and the quark one in particular.  

Finally, if the test quark is put far from the bulk of the wall, the string
will originate from this point making its way toward the wall, and will be 
screened inside the double Higgs region. 

However, if in the $\theta\neq \pi$ - vacuum 
the finite size ball surrounding a quark and consisting of any 
mixture of dyons and antidyons of any possible kind is excited, it will be 
unable to screen the quark charge as there is no border at infinity where the
residual polarization charge will be pushed out.

We have to make a reservation about the above described picture. As was
pointed out above (see footnote 6 and eq.(16)), 
there are reasons to expect that
the point $\theta=\pi$ is very especial for $SU(N_c=2),$ because the
vacuum energy density is likely to be exactly zero here. In this case, it
is natural if the dyon condensate also approaches zero therein. The theory 
is expected then to have massless dyons, etc. Clearly, there will be 
no confinement in this phase.  

\hspace{.5cm}{\bf 4.}\hspace{.5cm}
Let us point out now that the assumption about the confinement property
(at $\theta\neq \pi$) of the $SU(N_c)$ YM theory is not a pure guess, as
the above discussed nonanalytical (i.e. glued) structure of the vacuum
energy density, ${\ov E}_{vac}(\theta)$, is a clear signal about a phase
transition at some finite temperature. Indeed, at high temperatures the $
\theta$ - dependence of the free energy density is under control and is\,:$\,
\sim T^4(\Lambda/T)^{N_c b_o}\cos (\theta)$, due to a rare gas of instantons.
It is important for us here that it is perfectly analytic in $\theta$, and 
that this $\sim \cos(\theta)$ - behaviour is T-independent, i.e. it persists 
when the temperature decreases. On the opposite side, at $T=0$, the $\theta$ -
dependence is nonanalytic and, clearly, this nonanalyticity survives at
small temperatures. So, there should be a phase transition (confinement -
deconfinement) at some critical temperature, $T_c\simeq \Lambda$, where the 
$\theta$ - dependence changes qualitatively.  

Finally, let us show that, supposing that the monopoles indeed condense, we
can find out the form of the monopole field potential. So ($N_c=2$ and
$\theta=0$ for simplicity), let us return to the original partition
function, eq.(1), and let us suppose that we have integrated out the gluon
fields with two fields fixed: this time the field S and the monopole 
fields $M$ and $\ov M$. The potential will have the form:
\bq
U(S,M,{\ov M})=b_o S\ln\left (\frac{S}{\Lambda^4}\right )-
S f\left (z=\frac{{\ov M}M}{S^{1/2}}\right )\,,
\eq
and the stationary point equation is: 
$f^{\prime}(z_o)=0$. By the above assumption,
the function $f(z)$ is such that this equation has a nontrivial solution:
$({\ov M}M)=z_o S^{1/2}$, i.e. with $z_o\neq 0$. Substituting it now 
back to eq.(19), we obtain the monopole field potential:
\bq
U(M,{\ov M})=\frac{2\, b_o}{z_o^2}\left ({\ov M}M \right )^2\ln
\left (\frac{{\ov M}M}{\Lambda^2_M}\right ), 
\quad \Lambda_M^2=\Lambda^2 z_o e^{-f(z_o)/2 b_o}.
\eq
We see that the assumption made is selfconsistent, i.e. the monopole field
indeed condenses. It is sufficient to supply this potential with the 
simplest kinetic terms of the monopole and the (dual) neutral gluon fields,
to obtain the explicit solution for the electric string.

\begin{center}{\bf Acknowledgements}\end{center}

This work was supported in part by the Cariplo Foundation for Scientific
Research in collaboration with Landau Network-Centro Volta. I am deeply
indebted the whole staff, and especially to A.Auguadro, G.Marchesini and 
M.Martellini, for a kind hospitality extended to me during my stay at Milano.

The work is supported in part by RFFR, \# 96-02-19299-a.

\vspace{1cm}
\hspace{2cm} {\bf Appendix A}\\ 

The purpose of this appendix is to present in the explicit form the
asymptotic behaviour, at $\theta\ra \pm i\infty$, of the vacuum energy 
density of the $CP^N$-model (the leading term at $N\gg 1$).

The partition function can be written (in Euclidean space) in the form:
$$Z=\int d n d{\bar n}dA_{\mu}d\lambda \,\exp \left(- \int dx I(x) \right ),\,
I=\left\{{\overline{D_\mu n}}
\,D_\mu n-U{\bar n}n+\frac{N}{f_o}U-\frac{\th}{2\pi}F\right\},$$
where:
$$ D_\mu=\partial_\mu+iA_\mu,\,\, Q=\frac{F}{2\pi}=\frac{1}{2\pi} 
\epsilon_{\mu\nu} \partial_{\mu} A_{\nu},
\,\, \frac{-1}{N}T_{\mu\mu}=\frac{1}{2\pi}U=
\frac{f_o}{2\pi N} \overline{D_\mu n}\, D_\mu n, $$
 $U=-i\lambda,\, \th=i\theta,$\, and $f_o$ is the bare coupling: $f_o^{-1}=
(b_o/4\pi)\ln(M_o^2/\Lambda^2),\,b_o=1.$ Integrating out the n-field, one 
obtains the action:
$$\frac{1}{N}I=Tr\ln\left (-D^2_\mu-U\right )+\frac{U}{4\pi}\ln
\left (\frac{M_o^2}{\Lambda^2}\right )-
\frac{\tilde \theta}{N}\frac{F}{2\pi}\,.$$
As we need the potential only, the fields $U$ and $F$, which are direct 
analogs of the S and P fields in the YM theory, can be considered
as constant ones. The determinant for this case was calculated by F. Riva 
\cite{Riva}:
$$Tr\ln\left (-D^2_\mu-U\right )=\frac{-1}{4\pi}\int^\infty_{1/{M_o^2}}
\frac{dt}{t}\left [\frac{F}{\sinh (Ft)}\exp\{ Ut \}-\frac{1}{t}\right ].$$
The stationary point equations are:
$$\ln\left (\frac{M_o^2}{\Lambda^2}\right )=\int^\infty_{1/{M_o^2}}dt\,
\exp \{{\overline U}t \}\, \frac{\overline F}{\sinh ({\overline F}t)},\, 
\hspace{2cm} (a1)$$

$$\int^\infty_0\frac{dz}{z\sinh (z)} \exp\{\frac{{\overline U}}
{\overline F}\,z\}\left [ z\coth (z)-1\right ]=
2\,\frac{\tilde \theta}{N}\,. \hspace{1.cm} (a2) $$
We obtain from (a2) at $\th\ra \infty$:
$$\frac{{\overline U}(\th)}{{\overline F}(\th)}= 1-\Delta(\th),\,\,\,\, 
\Delta(\th)=\frac{N}{\th}\left [1-\frac{N}{\th}\ln\frac{\th}{N}+
O(\frac{N}{\th})\right ]\,, $$
and from (a1):
$$\ln\left (\frac{M_o^2}{\Lambda^2}\right )=\ln\left (\frac{M_o^2}
{\overline F(\th)}\right )+\frac{2}{\Delta(\th)}+O(1); \,\,\,
{\overline F}(\th)\ra const\left 
[\frac{\tilde \theta}{N}\exp \left \{\frac{\tilde \theta}{N}\right \} 
\right ]^{d/b_o}\,, $$
where d=2 is the space-time dimension, and $b_o=1$.

So, the behaviour of the vacuum energy density, 
$(1/2)T_{\mu\mu}=-N{\overline U}(\th)/4\pi$,
is nontrivial. ${\overline U}(\th)$ starts with the negative value $\sim
(-\Lambda^2)$ at $\th=0$, then increases monotonically with increasing $\th$,
passes zero at some $\th_o$ and becomes positive and large, approaching 
${\overline F}(\th)$ from below at large $\th$.


\begin{thebibliography}{99}

\bibitem{Sav}
G.K. Savvidy, Phys. Lett. {\bf B71} (1977) 133

\bibitem{II} V. Chernyak, BINP-98-62, hep-th/9808093
\bibitem{Shifm}
M. Shifman, Phys. Rep. {\bf 209} (1991) 341
\bibitem{ins}
V.A.Fateev, I.V.Frolov and A.S.Schwarz, Sov.J.Nucl.Phys. {\bf 30}(1979)590\\
B.Berg and M.Luscher, Commun.Math.Phys. {\bf 69} (1979) 57\\
R.D.Carlitz and D.A.Nicole, Nucl.Phys. {\bf B243} (1984) 307
\bibitem{Witten}
E. Witten, Phys. Lett.  {\bf B86} (1979) 283
\bibitem{Hooft}
G. 't Hooft, Nucl. Phys. {\bf B190 [FS3]} (1981) 455
\bibitem{Riva}
F. Riva, Nuovo Cim. {\bf A61} (1981) 69

\end{thebibliography}
\end{document}